\documentstyle[psfig]{mn}
\def\H0{{\it H}$_0$}

\def\q0{{\it q}$_0$}
\def\kmps{km~s$^{-1}$}

\def\ergps{erg~s$^{-1}$}

\def\nH{$N_{\rm H}$} 
\def\psqcm{cm$^{-2}$}
\def\ergpspsqcm{erg~cm$^{-2}$~s$^{-1}$}

\def\ltsima{$\; \buildrel < \over \sim \;$}
\def\simlt{\lower.5ex\hbox{\ltsima}}
\def\gtsima{$\; \buildrel > \over \sim \;$}
\def\simgt{\lower.5ex\hbox{\gtsima}}

\def\AX{AX\thinspace J1749+684\ }
\def\AXc{AX\thinspace J1749+684}

\title[X-ray absorption in Mrk~507]
{X-ray absorption in the strong FeII narrow-line Seyfert 1
galaxy Markarian\thinspace 507}
\author[K. Iwasawa, W.N. Brandt \& A.C. Fabian]
{\parbox[]{6.5in}{K. Iwasawa, W.N. Brandt$^{\star}$ and A.C. Fabian}\\
\\
Institute of Astronomy, Madingley Road, Cambridge CB3 0HA\\
$^{\star}$ Present address: The Pennsylvania State Univeristy, Department of Astronomy and Astrophysics, 525 Davey Park PA 16802, USA\\} 

\date{}

\begin{document}

\maketitle

\begin{abstract}
We present results from spectral analysis of ASCA data on the strong
Fe{\sc ii} narrow-line Seyfert 1 galaxy Mrk\thinspace 507. This galaxy
was found to have an exceptionally flat ROSAT spectrum among the
narrow-line Seyfert 1 galaxies (NLS1s) studied by Boller, Brandt \&
Fink (1996). The ASCA spectrum however shows a clear absorption
feature in the energy band below 2 keV, which partly accounts for the
flat spectrum observed with the ROSAT PSPC. Such absorption is rarely
observed in other NLS1s. The absorption is mainly due to cold (neutral
or slightly ionized) gas with a column density of (2--3)$\times
10^{21}$\psqcm. A reanalysis of the PSPC data shows that an
extrapolation of the best-fit model for the ASCA spectrum
underpredicts the X-ray emission observed with the PSPC below 0.4 keV
if the absorber is neutral, which indicates that the absorber is
slightly ionized, covers only part of the central source, or there is
extra soft thermal emission from an extended region. There is also
evidence that the X-ray absorption is complex; an additional edge
feature marginally detected at 0.84 keV suggests the presence of an
additional high ionization absorber which imposes a strong O{\sc viii}
edge on the spectrum. After correction for the absorption, the photon
index of the intrinsic continuum, $\Gamma\simeq 1.8$, obtained from
the ASCA data is quite similar to that of ordinary Seyfert 1
galaxies. Mrk\thinspace 507 still has one of the flattest continuum
slopes among NLS1, but is no longer exceptional. The strong optical
Fe{\sc ii} emission remains unusual in the light of the correlation
between Fe{\sc ii} strengths and steepness of soft X-ray slope.
\end{abstract}


\section{INTRODUCTION}

Mrk~507 is a narrow-line Seyfert galaxy at a redshift of $z = 0.0559$
(Halpern \& Oke 1987). Although this object had been classified as a
Seyfert-2 galaxy or a LINER based on the permitted line width
(FWHM$\sim 800$\kmps; Koski 1978) and line ratios (Heckman 1980), the
detection of strong optical Fe{\sc ii} permitted lines and a small
[OIII]$\lambda 5007$/H$\beta$ ratio favours classification as a narrow
line Seyfert 1 galaxy (NLS1; Halpern \& Oke 1987). NLS1 were first
studied by Osterbrock \& Pogge (1985); they have optical spectra which
are similar to those of normal Seyfert 1 galaxies but the Balmer lines
are narrow, typically with FWHM$\leq 2000$\kmps. A prototype of this
class of galaxy is I\thinspace Zw\thinspace 1 (see Goodrich 1989 for
the classification criteria in optical emission-line spectra for
NLS1).

NLS1s are different from ordinary Seyfert galaxies in their soft X-ray
properties, as shown by ROSAT observations. Rapid, large amplitude X-ray
variability and the lack of evidence for internal X-ray absorption
strongly suggest that we are seeing direct radiation from an unobscured
central source. This is consistent with the large X-ray luminosities of NLS1s
as previously pointed out by Halpern \& Oke (1987).
Another remarkable X-ray property is their steep soft
X-ray spectra. As shown in the ROSAT 0.1--2.4 keV
photon-index--FWHM(H$\beta$) diagram of Boller, Brandt \& Fink (1996,
hereafter BBF), photon indices ranging from 3.5 -- 5 are only found in
NLS1. Although the ROSAT photon indices for NLS1s are spread over a wide
range ($\Gamma = $ 2--5), the mean value ($\Gamma_{\rm NLS1}\simeq 3.1$)
is significantly larger than that ($\Gamma_{\rm Sy1}\simeq 2.3$) for
ordinary Seyfert 1 galaxies with broad Balmer lines (BBF). This may be
related to an intrinsic difference in the emission mechanism.

Mrk~507, which is an optically-classified NLS1, stands out since it shows
an exceptionally flat ROSAT spectrum for any type of Seyfert 1 galaxy ($\Gamma
= 1.6\pm 0.4$, BBF). Although possible X-ray flux variations were
detected from the ROSAT observations, the flat ROSAT spectrum does not
fit the general properties of NLS1s (BBF). The best-fit value of the
photon index is flatter by $\sim 0.7$ than even the mean ROSAT value for
ordinary Seyfert 1 galaxies. This could be taken to suggest that the spectral
slopes of NLS1s vary widely between individual objects, in contrast to
those of ordinary Seyfert 1 galaxies, which are found in a relatively
tight range. However, it is also possible that the flat spectrum is due
to a complicated absorption effect which cannot be resolved by the ROSAT
PSPC because of its poor spectral resolution and limited energy range.
Indeed, a strong anti-correlation 
between ROSAT spectral slope and FWHM(H$\beta$) is found 
in a sample of quasars studied by Laor et al (1997), 
which are presumably less absorbed than Seyfert galaxies, i.e., 
it appears that no flat X-ray spectrum quasars with narrow Balmer lines 
are found.

Since a large fraction of Seyfert 1 galaxies has been known to show
evidence for absorption by partially-ionized gas, the so-called 
``warm absorber'' (Halpern 1984) from a systematic study of ASCA
X-ray spectra (e.g., Reynolds 1997), similar absorption effects
can be expected in X-ray spectra of NLS1s.
Unlike the absorption by cold material found in many Seyfert 2 galaxies 
due to the obscuring torus, 
X-rays are absorbed mainly by partially-ionized oxygen in the
energy range between 0.5--2 keV in the case of a warm absorber.
The energy where the deepest feature appears depends on the ionization 
state of the absorber.

Our ASCA observation was aimed at investigating whether the remarkably
flat ROSAT spectrum of Mrk\thinspace 507 is the result of such
absorption effects or if indeed it has an intrinsically flat X-ray
continuum. Previous hard X-ray obsevations of Mrk\thinspace 507 with
non-imaging collimated instruments, such as the Ginga LAC, were
confused by the brighter Seyfert 1 galaxy Kaz\thinspace 163 which is
$\sim 10$ arcmin SW of Mrk\thinspace 507, so no spectral study of
Mrk~507 in the energy band above 2 keV has ever been made.

A value of the Hubble parameter of $H_0=50$ km s$^{-1}$ Mpc$^{-1}$ and a
cosmological deceleration parameter of $q_0={1\over 2}$ have been assumed
throughout.

\section{OBSERVATIONS AND DATA REDUCTION}

\begin{figure*}
\centerline{\psfig{figure=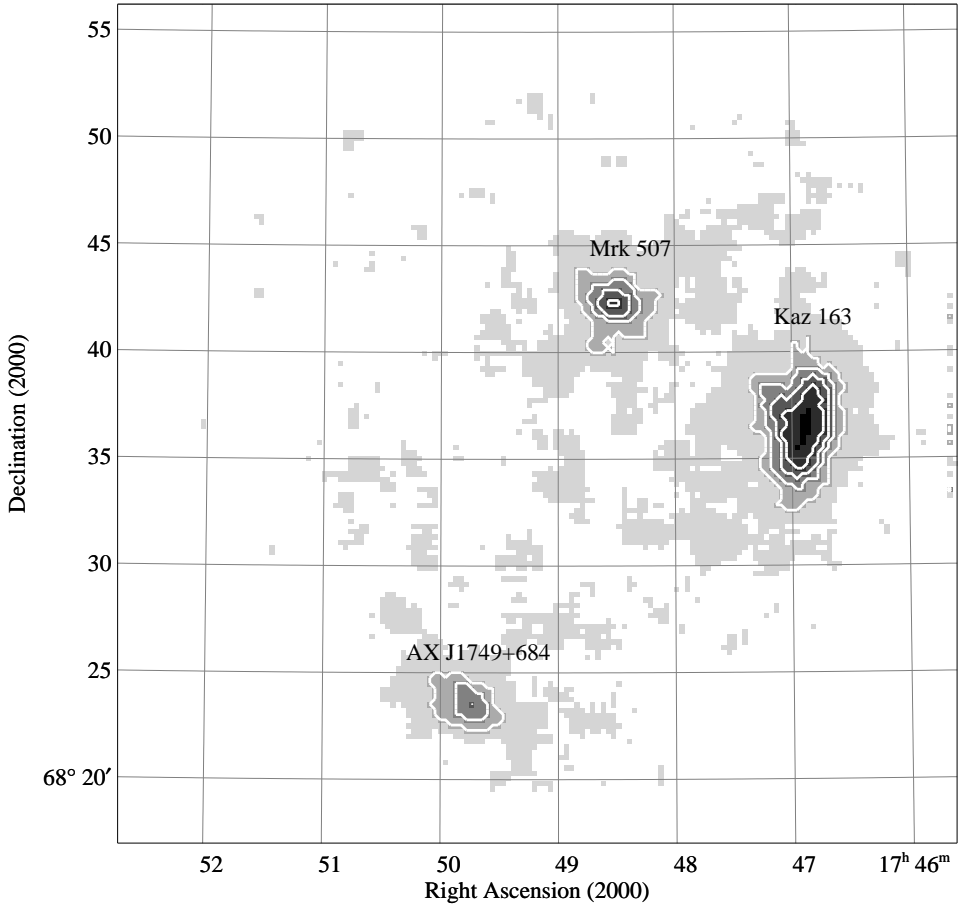,width=0.75\textwidth,angle=0}}
\caption{The ASCA GIS image (0.7--10 keV band) 
of Mrk~507, obtained by smoothing 
a summed G2+G3 image. Mrk~507, Kaz~163 and \AX are labelled.
Contours are at 33.3, 50.0, 66.6 and 83.3 per cent of the maximum
pixel value (see the text for source normalizations).}
\end{figure*}

Mrk~507 was observed by ASCA on 1995 December 16.  The Solid-state
Imaging Spectrometers (SIS) were operated in 2CCD/Faint mode.  The
lower level discriminator for the SIS was set at 0.47 keV to
prevent the telemetry from being saturated by the hot pixels of the
CCDs. This enabled us to use Faint mode throughout the observation.
In both SIS (S0 and S1), the most well-calibrated CCD chip was used
for Mrk~507, and the other was used for the nearby Seyfert 1 galaxy
Kaz~163 ($z=0.063$).  Kaz~163 was at the
edge of its chip with an off-axis angle $\sim 10$ arcmin.  About half
of the photons for Kaz~163 collected by the X-ray telescope were lost
due to this pointing configuration, which was the best possible given
restrictions on satellite operation.  The two objects were well within
the fields of view of the Gas Imaging Spectrometers (GIS; G2 and G3).
The net exposure time was about 37.9 ks for the SIS and 34.8 ks for
the GIS. The data reduction and analysis were performed
using {\sc ftools} and {\sc xspec} provided 
by the ASCA Guest Observer Facility at Goddard Space Flight Center. 
Spectral data for each source were taken from circular
regions with radii of 4 arcmin for the SIS and 5 arcmin for the GIS,
centred on each X-ray peak.  Background data for the SIS were taken
from local source-free regions on the same CCD chip for each source.
In addition to the two Seyfert galaxies, a previously unidentified
X-ray source was serendipitously detected by the GIS (see Fig. 1).  We
shall refer to this source as \AXc, and give details on this
object, including an optical follow-up observation, in a separate paper
(Iwasawa et al 1997a).
Background-subtracted count rates observed in each detector are shown
in Table 1. In spectral fits, we use response matrices generated by
SISRMG in FTOOLS for the SIS and the standard ones (version 4.0)
provided by the GIS team for the GIS.

These three objects were also observed in a ROSAT PSPC 
pointing at Mrk~507 on 1993 August 8--10.
The total exposure time for this pointing was 24.7 ks.
Results on Mrk~507 (BBF) and 
Kaz~163 (Brandt et~al. 1994) have been published. 
The public ROSAT data are reanalyzed here to obtain a consistent 
solution of the
spectrum of Mrk~507.

For the results of spectral fits, we quote 90 per cent confidence errors for
one parameter of interest.

\begin{table}
\begin{center}
\caption{Observed count rates in each detector for Mrk~507, Kaz~163
and \AXc. The integration times are 37.9 ks for the SIS and 34.8 ks
for the GIS. \AX was located outside of the SIS field of view. No
corrections for either XRT response or the vignetting of the detector
have been made.}
\begin{tabular}{lcccc}
Object & S0 & S1 & G2 & G3 \\
&\multicolumn{4}{c}{$10^{-2}$\thinspace counts\thinspace s$^{-1}$}\\
Mrk~507 & 1.65 & 1.26 & 1.16 & 1.62\\
Kaz~163 & 2.54 & 3.25 & 2.40 & 3.87 \\
\AX     & --- & --- & 0.99 & 0.77 \\
\end{tabular}
\end{center}
\end{table}

\section{RESULTS}

As Mrk\thinspace 507 is faint, the SIS data above 7 keV are unusable
for spectral analysis. In order to avoid uncertainties in response
matrices due to the level discriminator being set at 0.47 keV, we discard
the SIS data below 0.6 keV.  Spectral fitting was thus performed with
the 0.6--7 keV SIS data and the 0.9--10 keV GIS data.

\begin{figure}
\centerline{\psfig{figure=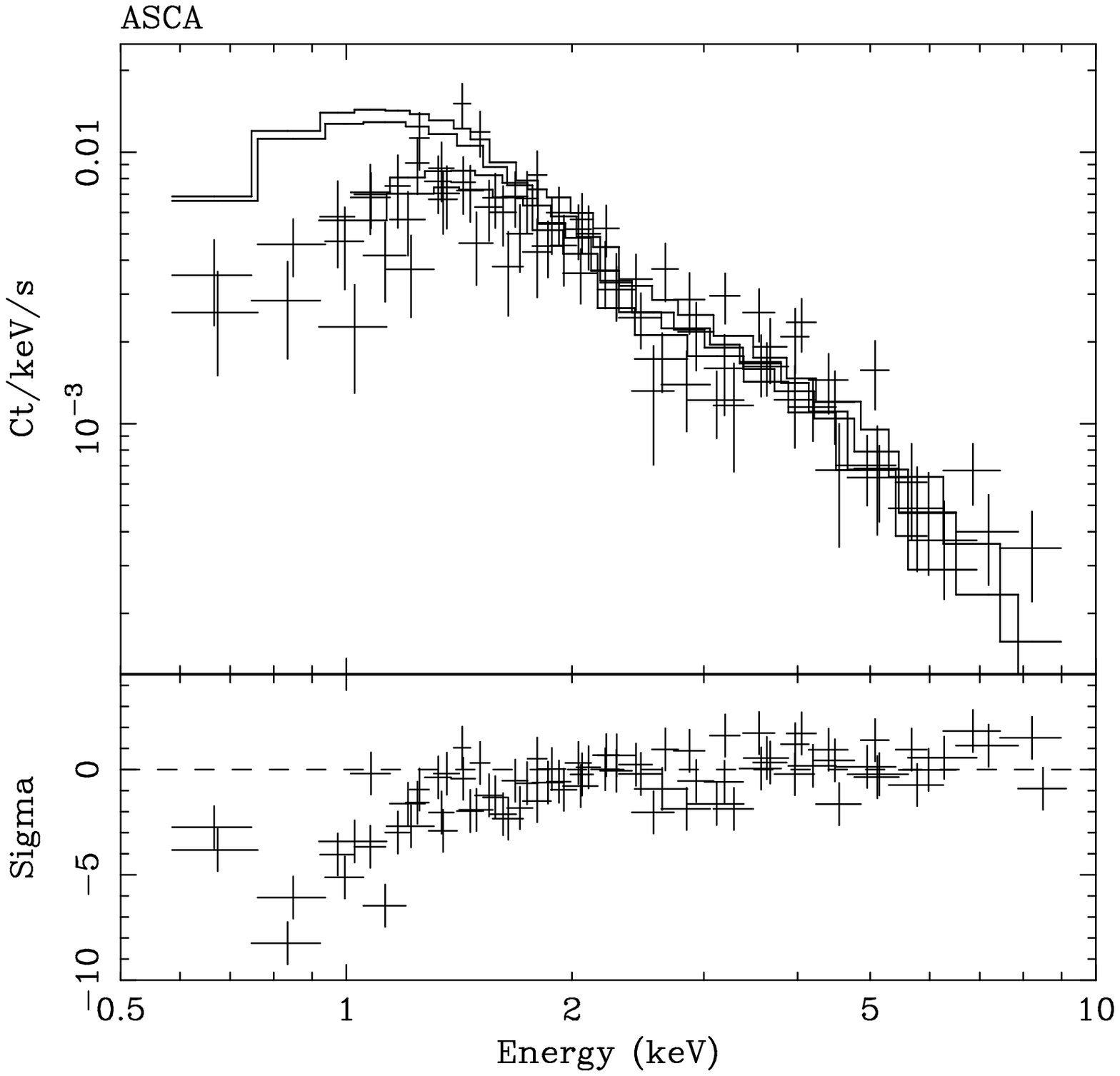,width=0.7\textwidth,angle=270}}
\caption{The ASCA spectra of Mrk\thinspace 507 from four detectors.
The solid line shows the best-fit power-law model with Galactic
absorption fitted to the 2--9 keV data. An absorption feature is
clearly seen below 1.5 keV in the residual plot.}
\end{figure}


Fitting with a power-law to the data 
above 2 keV from all four detectors gives a marginally acceptable fit with 
$\chi^2 = 94.76$ for 78 degrees of freedom. 
This yields a photon index 
$\Gamma = 1.7^{+0.2}_{-0.3}$ and a loose constraint on absorption, 
\nH $<1\times 10^{22}$\psqcm.
An extrapolation of the power-law model with Galactic absorption,
\nH $= 4.3\times 10^{20}$\psqcm ~(Stark et al 1992), 
below 2 keV shows a clear deficit of soft X-ray photons, probably 
due to absorption (Fig. 2).

When a power-law modified by cold absorption is fitted to the whole 
energy range of the ASCA data, the implied column density of 
cold material is \nH $\approx 3\times 10^{21}$\psqcm 
~and photon index $\Gamma\approx 1.7$ (see Table 2).
This fit however 
still leaves an excess below 0.9 keV,
a characteristic feature when warm absorption is relevant (see Reynolds 1996).
The inclusion of an additional absorption edge improves the quality of the fit
by $\Delta\chi^2 =6.9$ to $\chi^2=155.1$ for 152 degrees of freedom 
(see Table 2).
The edge threshould energy and optical depth are $E_{\rm th}=0.84\pm 0.07$ keV
and $\tau = 1.1\pm 0.6$, respectively. The significance of the edge is 
about at the 95 per cent confidence level, 
according to the F-test (Bevington 1969).
The edge feature is probably a mixture of edges due to O{\sc vii} (0.74 keV) 
and O{\sc viii} (0.87 keV),
as seen in Seyfert 1 galaxies, although our data do not have sufficient 
signal-to-noise to resolve them.
The relatively high edge energy suggests that O{\sc viii} could give a stronger
absorption edge than O{\sc vii} does.
The photon index is $\Gamma = 1.84\pm 0.2$, consistent with that obtained from
a simple power-law fit to the 2--10 keV data.
Significant cold absorption, 
\nH $= 2.0^{+1.4}_{-0.7}\times 10^{21}$\psqcm, is still required.

We also tried a partially covered power-law model.
Compared to the uniformly-absorbed power-law model with
warm absorber described above, the quality of the fit from this model is 
worse, and the improvement from the uniform absorber model is 
less significant (see Table 2).
Therefore the best description for the absorption in the ASCA spectrum is 
the cold absorber with \nH $\simeq 2\times 10^{21}$\psqcm ~plus 
a warm absorber which imposes the ionized oxygen edge around 0.8 keV.

No significant iron K$\alpha$ line at 6.4 keV is detected. 
Fitting the 2--10 keV data
with a power-law plus a gaussian line width $\sigma = 0.4$ keV, 
the mean iron K$\alpha$ line width of ASCA Seyfert 1 galaxy samples 
(Nandra et al 1996; Reynolds 1997), gives a 90 per cent upper limit on EW
of 730 eV.

The observed ASCA fluxes in the 0.6--2 keV and 2--10 keV band are 
$f_{\rm 0.6-2keV}=1.6\times 10^{-13}$\ergpspsqcm
~and $f_{\rm 2-10keV}=5.1\times 10^{-13}$\ergpspsqcm, respectively.
The absorption-corrected luminosities in the same energy bands are
$L_{\rm 0.6-2keV}=4.5\times 10^{42}$\ergps
~and $L_{\rm 2-10keV}=7.2\times 10^{42}$\ergps.

\begin{table*}
\caption{Spectral fitting results for the ASCA data. Four detectors are fitted
jointly. The Galactic absorption, \nH $= 4.3\times 10^{20}$\psqcm, is included
in all fits. PL is a power-law model. The edge threshould energy is 
corrected for redshift $z = 0.056$. $\ast$ A cold absorber which covers 
the central source partially; $\dag$ A covering fraction of the absorber.}
\begin{center}
\begin{tabular}{lccc}
\multicolumn{4}{c}{ASCA 2--9 keV}\\[5pt]
Model & $\Gamma$ & \nH & $\chi^2$/dof \\
& & $10^{21}$\psqcm & \\[5pt]
PL & $1.7^{+0.4}_{-0.3}$ & $<10$ & 95.11/77 \\
\end{tabular}
\end{center}
\vspace{3mm}
\begin{center}
\begin{tabular}{lcccccc}
\multicolumn{6}{c}{Whole ASCA band (0.6--9 keV)}\\[5pt]
Model &  $\Gamma$ & \nH & $f_{\rm C}^{\dag}$ & $E_{\rm th}$ & 
$\tau$ & $\chi^2$/dof \\
&& $10^{21}$\psqcm && keV && \\[5pt]
PL  & $1.76^{+0.22}_{-0.20}$ & $3.4^{+1.5}_{-1.3}$ & 1.0 &
--- & --- & 162.0/154 \\
PL + Edge & $1.84^{+0.21}_{-0.19}$ & $2.0^{+1.4}_{-0.7}$ & 1.0 & 
$0.84\pm 0.07$ & $1.1\pm 0.6$ & 155.1/152 \\
PL + PC$^{\ast}$ & $1.92^{+0.31}_{-0.26}$ & $6.0^{+3.8}_{-3.7}$ &
0.79 ($\geq 0.64$) & --- & --- & 159.7/153 \\
\end{tabular}
\end{center}
\end{table*}

We now consider the reason for the flat ROSAT spectrum.
A simple power-law model gives a photon index $\Gamma = 1.6\pm 0.4$
with no excess absorption above Galactic value (Fig. 3a and BBF).
It is however clear from the ASCA spectrum (see Fig. 2) 
that almost the entire ROSAT band ($\leq 2$ keV) is affected by absorption.
As no absorption-free continuum is seen by ROSAT, 
a simple power-law fit is misleading. 
The absorption detected by ASCA is a likely explanation for the flat
ROSAT spectrum.

\begin{figure}
\centerline{\psfig{figure=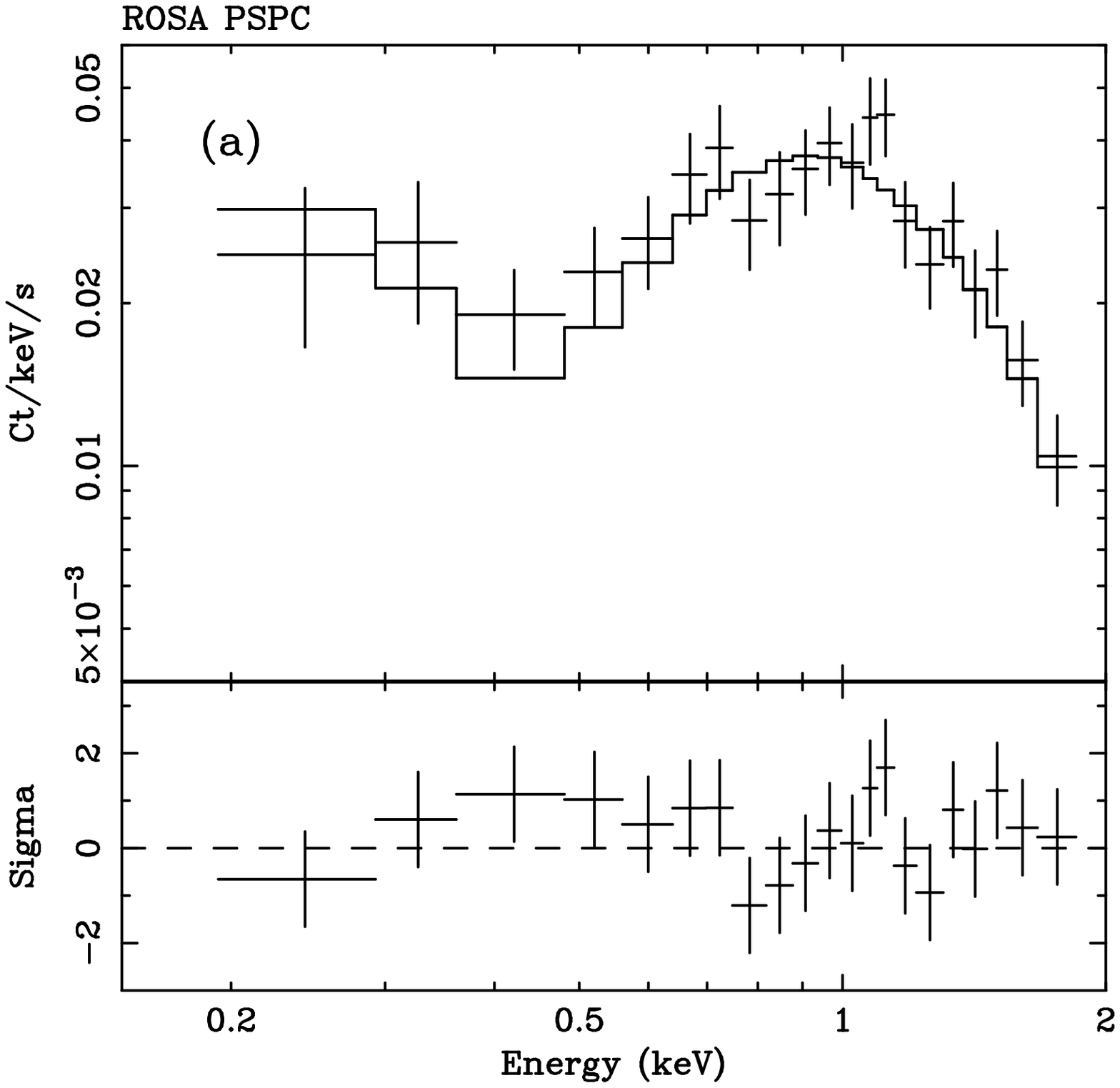,width=0.7\textwidth,angle=270}}
\vspace{-1cm}
\centerline{\psfig{figure=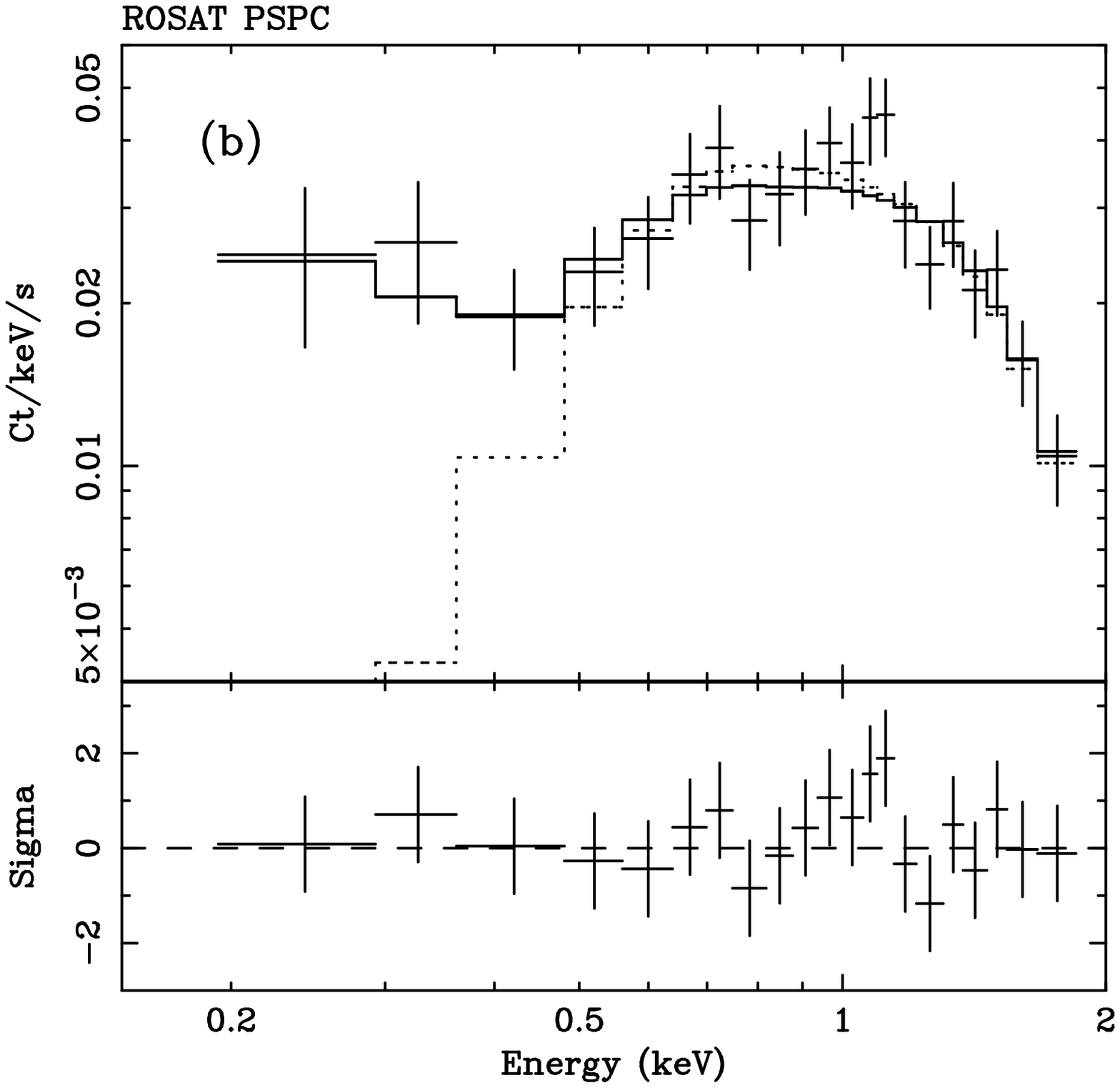,width=0.7\textwidth,angle=270}}
\caption{The ROSAT PSPC spectrum of Mrk\thinspace 507 (BBF). (a) A
simple power-law fit with $\Gamma\approx 1.6$ and \nH $= 4\times
10^{20}$\psqcm. (b) A power-law with $\Gamma = 2.4$ modified by the
0.8 keV edge and cold absorption (\nH $\approx 2\times 10^{21}$\psqcm;
dotted line) or ionized absorption ($N_{\rm W}\approx 3\times
10^{21}$\psqcm ~and $\xi\approx 2$; solid line).  The residual plot is
for the ionized absorber fit.}
\end{figure}

It should be noted that calibration uncertainties between ASCA and
ROSAT currently remain. There is significant disagreement between ASCA
and ROSAT spectral slopes in active galaxies (e.g., Yaqoob et al 1994;
Fabian et al 1995); ROSAT spectra, especially ones taken after the
gain change of the PSPC in 1991, often give steeper photon indices
than ASCA spectra (by $\Delta\Gamma\sim$0.5, e.g., ASCA GOF
1995).
Actually this discrepancy between photon indices from the two
satellites is confirmed by observations of the normal Seyfert 1
galaxy, Kaz\thinspace 163, in the same field of view as Mrk\thinspace
507. This shows a featureless power-law spectrum. A similar difference
($\Delta\Gamma\simeq 0.5$) between the ROSAT photon index ($\Gamma =
2.5\pm 0.1$, Brandt et al 1994) and the ASCA one ($\Gamma = 2.0\pm
0.1$) is found (although the two observations were not
simultaneous). A comparison of Einstein Observatory IPC and ROSAT PSPC
data of the EMSS AGN also gives a similar trend ($\Delta\Gamma$ =
0.4--0.5, Ciliegi \& Maccacaro 1996). The steepness of ROSAT spectra
has often been reconciled by assuming strong soft excess emission in
the ROSAT band. However, this is not the only reason and a
cross-calibration error seems to also be important. Even in the 0.5--2
keV band where both the PSPC and the ASCA SIS are sensitive, a
significant difference in spectral slope was found in simultaneous
data on NGC5548 (Fabian et al 1994) and also for Kaz\thinspace 163. A
consistency in spectral slope measurement between ASCA and previous
instruments such as the Einstein IPC, EXOSAT ME/LE and Ginga LAC is
confirmed by observations of the Galactic synchrotron nebula 3C58
(Dotani et al 1996). Although uncertainties in response matrices for
the SIS at energies around 0.5 keV, which could cause an error in
absorption measurements by $\Delta$\nH $\sim 2\times 10^{20}$\psqcm,
are reported (Dotani et al 1996), this could not be a problem in our
analysis because we discarded the data below 0.6 keV. Therefore at the
current status of calibrations, $\Delta\Gamma\sim 0.5$ might be
appropriate when ASCA and ROSAT data are compared.

We therefore naively assume a ROSAT photon index of 2.4 
from the ASCA result ($\Gamma = 1.84\pm 0.2$) and examine the 
PSPC data.
It now fits well to the ROSAT data above 0.4 keV, when the power-law is
modified by cold absorption with \nH $= (1.7\pm 0.5)\times 10^{21}$\psqcm
~and an additional edge at the rest energy of 0.84 keV
and optical depth $\tau\simeq 0.6 (\leq 1.1)$, similar to the ASCA
spectral fits (see Table 2).
However, some excess emission is found above the model in the energy
band below 0.4 keV, as illustrated in Fig. 3b.
Cold absorption with the implied column density suppresses X-ray photons
of those energies.

The excess emission can be explained by any of three possibilities;
(1) replacing the cold absorber by a low ionization warm absorber; (2)
a partially covering absorber; and (3) extra emission from an extended
region.  In the first case, the absorber is slightly ionized rather
than neutral so that the spectrum would recover below 0.5 keV (see
Fig. 3b).  This absorber should be different from the one with the
ionization parameter, $\xi = L/(n^2R)\sim 10^2$ which imposes the
strong O{\sc viii} edge around 0.8 keV.  If a photoionized absorber
calculated by CLOUDY (Ferland 1996) is fitted, the best-fit parameters
are $\xi\simeq 2$ and column density $N_{\rm W}\simeq 3\times
10^{21}$\psqcm.  In the second case, some fraction of the central
source escape from the cold absorber. If a partial covering model is
fitted, the column density and covering fraction of the absorber are
\nH = $2^{+2}_{-1}\times 10^{21}$\psqcm ~and $f_{\rm C} = 0.75\pm
0.1$, respectively. Finally, in the third case, thermal emission with
a temperature of $10^7$ K and luminosity of $\sim 10^{42}$\ergps ~must
be associated with the galaxy. An extended optical emission line
region has been found in Mrk\thinspace 507 (Halpern \& Oke 1987). Many
nearby active galaxies having extended optical emission nebula,
especially Seyfert 2 galaxies, show thermal emission in the soft X-ray
band, although with much lower luminosities (e.g., Wilson, Elvis \&
Bland-Hawthorn 1992; Weaver et al 1994; Makishima et al 1994; Morse et
al 1995; Iwasawa et al 1997).

\section{DISCUSSION}

Intrinsic absorption has been found in the ASCA spectrum of this
narrow-line Seyfert 1 galaxy. It is likely that the absorption makes
the ROSAT spectrum appear flat. The absorption is mainly due to a
cold, or possibly low ionization, absorber with a column density of
(2--3)$\times 10^{21}$\psqcm. Soft X-ray emission is seen below 0.4
keV detected with the ROSAT PSPC, three options are possible (see
Section 3.1); (1) the absorber is slightly ionized ($\xi\sim 2$); (2)
the covering fraction of the absorber is less than unity ($\sim 75$
per cent); and (3) there is extra thermal emission with a temperature
of $\sim 10^7$ K. In any case, the excess soft X-ray emission,
escaping from the absorption detected with ASCA, may be related to
even lower energy emission claimed from Extreme Ultraviolet Expolorer
data (Fruscione 1996). Note that the ASCA spectrum is not sensitive to
the ionization of absorber (1), because the recovery of the spectrum
is out of the ASCA bandpass ($<0.6$ keV).  An additional edge
absorption feature at 0.84 keV is marginally detected (at $\sim$95 per
cent confidence level), implying another high ionization absorber
($\xi\sim 10^2$) which imposes a strong O{\sc viii} edge. Such a multi
warm-absorber is not unusual in Seyfert 1 galaxies, as found in
MCG--6-30-15 (Otani et al 1996), although the ionization parameters
are different in Mrk\thinspace 507. It should be noted that the cold
absorption detected from Mrk\thinspace 507 is unusual for NLS1s which
normally show no evidence for cold absorber although some may have
warm absorbers (e.g., Brandt et al 1994).

The intrinsic continuum slope of Mrk\thinspace 507 is found to be 
$\Gamma\approx 1.8$ (see Table 2), similar to that of 
ordinary Seyfert 1 galaxies.
It is in a good agreement with the mean value, $\Gamma = 1.81$ found for
an ASCA sample of 24 Seyfert 1 galaxies (Reynolds 1997).
This still does not fit the steep spectral nature of most of NLS1s,
and means that Mrk\thinspace 507 has the flattest photon index 
compared with other NLS1s with similar FWHM(H$\beta$) in the diagram of 
$\Gamma$--FWHM(H$\beta$) in BBF even after correction for absorption.
The X-ray absorption may imply obscuration of the broad-line region 
which results in narrow width of the Balmer lines.

Mrk\thinspace 507 is an extreme Fe{\sc ii} AGN (Fe{\sc
ii}/H$\beta\simeq 2.9$, Lipari 1994). The rather normal continuum
slope is then inconsistent with the correlation between soft X-ray
slope and strength of optical Fe{\sc ii} emission found from a study
of Einstein Observatory IPC data (Wilkes, Elvis \& McHardy 1987; see
also Boroson 1989).  A
possible difference between Mrk\thinspace 507 and steep-spectrum NLS1s
is its X-ray luminosity. The total X-ray luminosity of Mrk\thinspace
507 is about $1\times 10^{43}$\ergps ~whilst steep-spectrum NLS1s have
X-ray luminosities typically $10^{44}$\ergps ~or larger (e.g.,
$2\times 10^{44}$\ergps ~for I\thinspace Zw\thinspace 1, Halpern \&
Oke 1987; $3\times 10^{44}$\ergps ~for IRAS\thinspace 13224--3809,
Boller et al 1993).  There are other X-ray quiet AGNs with extreme
Fe{\sc ii}. Mrk\thinspace 231 (Rigopoulou et al 1996) and 
IRAS\thinspace 07598+6508
(Lawrence et al 1997) are typical of those.  These Fe{\sc ii} AGNs
including Mrk\thinspace 507 appear to be different from the powerful
steep-spectrum NLS1s, and the relationship between strength of optical
Fe{\sc ii} emission and soft X-ray properties remains uncertain.

The far-infrared peaked SED (e.g., Lipari 1994), strong optical Fe{\sc ii},
weak [OIII]$\lambda 5007$ (EW is about 7 \AA; 
J. Halpern, private communication) and evidence for significant 
intrinsic X-ray absorption seen in Mrk\thinspace 507 are also generally 
seen in broad absorption line quasars. 
It would be interesting to see
if Mrk 507 also shows broad aborption lines in the ultraviolet due to 
absorption by the same material that we see with ASCA.

\section{CONCLUSIONS}

We detected significant absorption in the ASCA spectrum of a NLS1,
Mrk\thinspace 507, which plausibly causes the apparently flat spectrum
in the ROSAT band measured by BBF. The absorber is neutral or possibly
slightly ionized, and the column density is 2--3$\times
10^{21}$\psqcm, depending on models. In addition to the cold absorber,
marginal evidence for an O{\sc viii} edge due to another warm absorber with
high ionization parameter was found. The spectral slope of the
intrinsic continuum is similar to ordinary Seyfert 1 galaxies rather
than NLS1s that generally have steeper soft X-ray continua. As one of
extreme Fe{\sc ii} AGNs (Lipari 1994), this is also unusual in the
light of the correlation between EW(Fe{\sc ii}) and steepness of soft
X-ray spectra claimed by Wilkes, Elvis \& McHardy (1987).

\section*{ACKNOWLEDGEMENTS}

We thank Jules Halpern for helpful discussions.
We thank all the members of the ASCA team who maintain the satellite and
carry out operations. ACF, KI and WNB thank the Royal Society,
the PPARC, the Smithsonian Institution, respectively, for support.

\bsp

\end{document}